\documentclass[12pt]{article}
\usepackage{fullpage,amsmath}
\usepackage{setspace, lscape}
\usepackage{graphicx, psfrag, amsfonts, bm, amssymb}
\usepackage{natbib}
\usepackage{wrapfig}
\usepackage{pdfsync}
\usepackage[urlcolor=blue]{hyperref}
\usepackage{multirow}
\usepackage{hhline}
\usepackage{caption}
\usepackage{url, xr}
\usepackage[table]{xcolor}
\usepackage{threeparttable}
\usepackage{graphicx}
\usepackage{adjustbox}
\usepackage{array}
\usepackage{lmodern}
\usepackage[utf8]{inputenc}
\usepackage[utf8]{inputenc}
\usepackage{hyperref}
\usepackage{amsmath, amsfonts, amssymb, bm, setspace, lscape, graphicx, psfrag, wrapfig, multirow, hhline, caption, url, xr, xcolor, multicol, booktabs, subcaption, ulem}
\usepackage{mathtools}

\begin{document}
	\doublespacing

	\title{BMW: Bayesian Model-Assisted Adaptive Phase II Clinical Trial Design for Win Ratio Statistic}

	\author{Di\ Zhu\\
		{\small Department of Biostatistics and Health Data Science;}\\
		{\small School of Medicine, Indiana University, Indianapolis, IN}\\	  		       
		Yong\ Zang$^{*}$\\
		{\small Department of Biostatistics and Health Data Science;}\\
		{\small Center for Computational Biology and Bioinformatics;}\\  
		{\small School of Medicine, Indiana University, Indianapolis, IN}\\	
	}

	\date{}
	\maketitle
	
	* Correspondence to:  Yong Zang, Ph.D., Department of Biostatistics and Health Data Science; Center for Computational Biology and Bioinformatics, School of Medicine, Indiana University, Indianapolis, IN (email: zangy@iu.edu).
	
	\newpage 
	 
	\begin{abstract}
	\noindent 	The win ratio (WR) statistic is increasingly used to evaluate treatment effects based on prioritized composite endpoints, yet existing Bayesian adaptive designs are not directly applicable because the WR is a summary statistic derived from pairwise comparisons and does not correspond to a unique data-generating mechanism. We propose a Bayesian model-assisted adaptive design for randomized phase II clinical trials based on the WR statistic, referred to as the BMW design. The proposed design uses the joint asymptotic distribution of WR test statistics across interim and final analyses to compute posterior probabilities without specifying the underlying outcome distribution. The BMW design allows flexible interim monitoring with early stopping for futility or superiority and is extended to jointly evaluate efficacy and toxicity using a graphical testing procedure that controls the family-wise error rate (FWER). Simulation studies demonstrate that the BMW design maintains valid type I error and FWER control, achieves power comparable to conventional methods, and substantially reduces expected sample size. An R Shiny application is provided to facilitate practical implementation. 
	\end{abstract}

\section{Introduction}  

Bayesian adaptive designs have gained substantial popularity in the development and conduct of early-phase clinical trials \citep{fda26}. Traditionally, these designs rely on parametric models to characterize the treatment–outcome relationship and to derive posterior probabilities from observed data, which are then used to guide patient allocation and treatment selection at interim and final analyses \citep{berry10}. However, such model-based designs can be difficult to understand, rely on strong parametric model assumptions that may not hold for real-world clinical data, and require substantial computational resources and expensive infrastructure for implementation in clinical practice. \citep{che12}.

In recent years, a new class of Bayesian adaptive designs, referred to as the model-assisted designs, has been proposed as an alternative to model-based designs \citep{yuan19}. Like model-based designs, model-assisted designs rely on statistical models to support decision making. However, their decision rules can be fully pre-specified prior to trial initiation and explicitly incorporated into the study protocol. In general, model-assisted designs often demonstrate superior operating characteristics compared with model-based designs and can be implemented more straightforwardly, often using freely available software \citep{yuan22}.  

In phase II clinical trials, the Bayesian optimal phase II (BOP2) design is a transparent, flexible, and efficient model-assisted trial design for evaluating treatment effect \citep{zhou17}. As a Bayesian adaptive design, BOP2 offers several desirable properties, including flexibility in the timing and frequency of interim analyses and the ability to stop a trial early for either futility or superiority \citep{xu25}. At the same time, similar to frequentist designs, BOP2 can strictly control the type I error rate and maximize statistical power through simulation-based numerical optimization of design parameters.  Although it was originally developed for single-arm trials with categorical outcomes, the BOP2 design has since been extended to randomized controlled trials (RCT) \citep{zhao22} and to time-to-event outcomes \citep{zhou20}.

Based on the Dirichlet–multinomial model, the BOP2 design can readily accommodate multiple types of outcomes within a unified multiple hypothesis testing framework. For example, consider a RCT with two efficacy endpoints: objective tumor response (OR) and 3-month event-free survival (EFS3). The BOP2 design can be used to formulate and test hypotheses such that the new treatment is declared successful if it demonstrates a statistically significant improvement over the control in either OR or EFS3. However, despite this flexibility, the BOP2 design treats multiple endpoints exclusively as co-primary and model the joint distribution for all the endpoints of interest, which may not fully reflect clinical practice, particularly with the increasing use of composite endpoint.

In clinical trials with multiple endpoints,  the composite endpoint is often used to combine multiple endpoints into a single index, thereby improving statistical efficiency, simplifying hypothesis testing, and avoiding complex joint modeling approaches \citep{free03}. However, the conventional definition of a composite endpoint treats all components as equally important and typically considers only the first event. As a result, it ignores clinical priorities and the timing of multiple endpoints, both of which are often critically important in clinical practice. To address these limitations, the win ratio (WR) statistic has been proposed as an alternative that incorporates a clinically meaningful hierarchy of endpoints through pairwise comparisons between treatment and control arms, sequentially prioritizing more important outcomes and yielding a more appropriate assessment of the overall treatment benefit \citep{po2012}.  

Since its introduction, the WR statistic has been used in many clinical trials to evaluate the overall treatment effect \citep{red20}. The statistical properties of the WR statistic have also been extensively studied in the literature, including large-sample inference approach developed under the U-statistic framework \citep{be16} and sample size formulas \citep{mao21, yu22}. However, little work has focused on Bayesian adaptive designs for the WR statistic, and existing methods such as the BOP2 design are not directly applicable in this setting. A key challenge is that the posterior probabilities used in the BOP2 design and other Bayesian adaptive designs are derived directly from the observed data, whereas the WR statistic is a summary measurement of pairwise comparison rather than the raw observed data itself. Moreover, the WR statistic is highly flexible and general, placing no restrictions on either the number of endpoints or the correlation structure among them. Consequently, theoretical speaking, infinitely many underlying data-generating mechanisms may yield the same WR value, which substantially complicates the simulation-based calibration required for any Bayesian adaptive designs.   

In this paper, we address this research gap by developing a Bayesian model-assisted design tailored to the WR statistic, referred to as the BMW design, for phase II RCT. To overcome the aforementioned challenges, we derive the asymptotic joint distribution of the WR statistic across different interim and final analysis time points. Then, based on this distribution, the posterior probability to evaluate the overall treatment effect based on the WR statistic is calculated to guide trial conduct. The proposed approach offers several advantages. First, it eliminates uncertainty arising from arbitrary assumptions about the underlying data-generating mechanisms. Second, for simulation-based calibration and design parameter optimization, the proposed framework enables direct simulation of the WR statistic, rather than repeatedly generating outcome data and recomputing the WR for each simulation replicate, thereby substantially improving computational efficiency.

The BMW design allows for early stopping for either futility or superiority and is further extended to accommodate both efficacy and toxicity outcomes through a graphical testing procedure. Similar to the BOP2 design, the BMW design strictly controls the type I error rate while maximizing power, and its stopping boundaries can be pre-tabulated prior to trial initiation. In addition, we provide freely available R Shiny Apps to facilitate its implementation in clinical practice.

\section{Method}	
We first consider the setting in which only efficacy outcomes are evaluated. Consider a phase II RCT that uses the WR statistic to quantify treatment effectiveness with multiple efficacy endpoints. Each patient in the trial is randomized to receive either the investigational treatment $s=1$ or the control treatment $s=0$ with the randomization ratio $\phi$ and $1-\phi$, respectively. The trial consists of $R-1$ interim analyses conducted when enrollment reaches $N_1, \ldots, N_{R-1}$, followed by a final analysis after all $N_R$ participants have been enrolled and evaluated.

We consider the unmatched setting in this paper, which is more applicable in practice. To calculate the WR statistic, each patient in the treatment group is paired with every patient in the control group. For each pairwise comparison, a winner and a loser are identified unless the outcome results in a tie. If a comparison is inconclusive (i.e., yields a tie), the next highest-priority endpoint is evaluated using the same procedure. This process continues sequentially through the predefined hierarchy of endpoints.

Let $p_{W}$, $p_{L}$ and $p_{T}$ denote the probabilities of win, loss and tie, and $\theta={\rm log}\big( \frac{p_W}{p_L}  \big)$ be the logarithm of the WR. We are interested in testing the following hypotheses for efficacy:
\begin{eqnarray}
	{\rm H}_{0,E}: \theta=0 \ \ \  \leftrightarrow \ \ \ {\rm H}_{1,E}: \theta>0. 
\end{eqnarray}	
During the $r$th analysis $(r=1, \cdots, R)$, among a total of $\phi(1-\phi)N_r^2$ comparisons, let $N_{r,W}$, $N_{r,L}$ and $N_{r,T}$ be the numbers of wins, losses and ties. Then, $p_{W}$, $p_{L}$ and $p_{T}$ can be estimated as $\widehat{p}_{r,W}=\frac{N_{r,W}}{\phi(1-\phi)N_r^2}$, $\widehat{p}_{r,L}=\frac{N_{r,L}}{\phi(1-\phi)N_r^2}$, $\widehat{p}_{r,T}=\frac{N_{r,T}}{\phi(1-\phi)N_r^2}$, and the WR statistic can be constructed as ${\rm WR}_{r}=\frac{\widehat{p}_{r,W}}{\widehat{p}_{r,L}}$. Subsequently, during the $r$th analysis, $\theta$ can be estimated as $\widehat{\theta}_r={\rm log}\big(  {\rm WR}_r   \big)={\rm log}\big(  \frac{N_{r,W}}{N_{r,L}}  \big)$. Following \cite{yu22}, the variance of $\widehat{\theta}_r$ can be approximated as ${\rm Var}( \widehat{\theta}_r  )\approx\frac{4(1+p_T)}{3\phi(1-\phi)(1-p_T)N_r}$. Then, the Wald-type win ratio test statistic under the $r$th analysis can be constructed as:

\begin{eqnarray}
	z_r=\frac{ {\rm log}\big(  {\rm WR}_r   \big)   }{ \sqrt{  \widehat{ {\rm Var}( \widehat{\theta}_r )   }} } =\frac{  {\rm log}\big(  \frac{N_{r,W}}{N_{r,L}}  \big)   }{  \sqrt{   \frac{4(1+\widehat{p}_T)}{3\phi(1-\phi)(1-\widehat{p}_T)N_r}    } }.
\end{eqnarray}

Let $R^{\dagger}$ denote the current analysis time, and let  $\mathbf{Z}_{R^{\dagger}} = (z_1, \ldots, z_{R^{\dagger}})^{\mathrm{T}}$ be the vector of the win ratio test statistic based on all the accumulated data. The following theorem summarizes the asymptotic 
distribution of $\mathbf{Z}_{R^{\dagger}} $, with the proof provided in the online supplementary material.

\vskip 0.2in
\noindent {\it Theorem 1.} Let ${\rm I}_r=\frac{1}{   {\rm Var}( \widehat{\theta}_r )}=\frac{3\phi(1-\phi)(1-p_T)N_r}{4(1+p_T)}$ be the Fisher information value under the $r$th analysis and denote  $\mathbf{Z_{R^{\dagger}}} = (z_1, \ldots, z_{R^{\dagger}})^{\mathrm{T}}$. Under the regularity conditions, $\mathbf{Z_{R^{\dagger}}}$ asymptotically follows the multivariate normal distribution:
\begin{eqnarray}
	&&\mathbf{Z_{R^{\dagger}}}\sim {\cal N}({\bf M_{R^{\dagger}}},  {\bf \Sigma_{ R^{\dagger} }   }), \ \ \, {\bf M_{ R^{\dagger}}}=(\mu_1, \cdots, \mu_{R^{\dagger}})^{\rm T}, \ \ \ {\bf \Sigma_{ R^{\dagger} }   }=\begin{pmatrix}
		1           & \rho_{1,2}  & \cdots & \rho_{1,R^{\dagger}} \\
		\rho_{2,1}  & 1           & \cdots & \rho_{2,R^{\dagger}} \\
		\vdots      & \vdots         & \ddots & \vdots     \\
		\rho_{R^{\dagger},1}  & \rho_{R^{\dagger},2}  & \cdots & 1
	\end{pmatrix}; \nonumber\\
	&&\mu_r=\theta \sqrt{ {\rm I}_r }, \ \ \  \rho_{r_1, r_2 }=\rho_{r_2, r_1}=\sqrt{   \frac{  {\rm I}_{r_1}   }{   {\rm I}_{r_2}  }  } \ {\rm for} \ r_1<r_2.  \nonumber
\end{eqnarray}
\vskip 0.2in

Let $\mathbf{Z_{R^{\dagger}}^{(o)}  }$ be  the observed value of $\mathbf{Z_{R^{\dagger}}  }$.  Based on Theorem 1, the likelihood function of $\mathbf{Z_{R^{\dagger}}^{(o)} }$ is expressed as:

\begin{eqnarray}
	{\rm L}(   \mathbf{Z_{  R^{\dagger}    }^{(o)}} \mid \theta   )=(2\pi)^{-R^{\dagger} /2}|  \Sigma_{R^{\dagger} }  |^{-1/2} {\rm exp}  \Big(  -\frac{1}{2}  ( \mathbf{Z_{R^{\dagger}}^{(o)} }-{\bf M_{R^{\dagger}}}  )^{ \rm T}  \Sigma_{R^{\dagger} }^{-1}   ( \mathbf{Z_{R^{\dagger}}^{(o)}}-{\bf M_{R^{\dagger}}}  )      \Big).
\end{eqnarray}	
After assigning $\theta$ a non-informative normal prior $\theta \sim {\cal N}(\theta_0=0, \sigma_0^2=100)$ with the density function defined by $f(\theta)$, the posterior distribution of $\theta$ can be derived from $ f(\theta \mid \mathbf{Z_{  R^{\dagger}    }^{(o)} } ) \propto {\rm L}(   \mathbf{Z_{  R^{\dagger}    }^{(o)}} \mid \theta   ) f(\theta) $. 	The following theorem provides the analytic posterior distribution of $\theta$, with the detailed calculation provided in the online supplementary material. 

\vskip 0.2in
\noindent {\it Theorem 2}. Let ${\bf B_{ R^{\dagger} }}=( \sqrt{I_1}, \cdots, \sqrt{ I_{ R^{\dagger}   }  }   )^{\rm T}$. The posterior distribution of $\theta$, $ \theta \mid {\bf Z_{ R^{\dagger}}^{(o)} }$, follows the normal distribution:
\begin{eqnarray}
	&&  \theta \mid {\bf Z_{ R^{\dagger}}^{(o)} } \sim {\cal N}(\theta_{R^{\dagger}}, \sigma_{R^{\dagger}}^2); \nonumber\\
	&& \sigma_{R^{\dagger}}^2=  \Big(   \frac{1}{\sigma_0^2}+  {\bf B_{ R^{\dagger} }^{\rm T}}  {\bf \Sigma_{ R^{\dagger} }^{-1}  }    {\bf B_{ R^{\dagger} }} \Big)^{-1}, \ \ \ \theta_{R^{\dagger}}=\sigma_1^2 \Big(   \frac{\theta_0}{\sigma_0^2}+  {\bf B_{ R^{\dagger} }^{\rm T}}  {\bf \Sigma_{ R^{\dagger} }^{-1}  }  {\bf Z_{R^{\dagger}}^{(o)}}       \Big). 
\end{eqnarray}	
\vskip 0.2in

Let $f(\theta \mid \mathbf{Z_{  R^{\dagger}    }^{(o)}} )$ be the density function for the posterior distribution of $\theta$. By Theorem 2, we can calculate the posterior probability (PP) for efficacy as:

\begin{eqnarray}
	{\rm PP}_{E, R^{\dagger}}={\rm Pr}( \theta>0 \mid \mathbf{Z_{  R^{\dagger}    }^{(o)}} )=\int_{0}^{+\infty} f(\theta \mid \mathbf{Z_{  R^{\dagger}    }^{(o)}} ) d\theta.
\end{eqnarray}	 
This posterior probability will be used to guide trial conduct. In particular, the BMW design with efficacy outcomes only is summarized as follows.

\begin{itemize}
	\item[1.1] Start the trial by enrolling and randomizing patients to receive either the investigational treatment $s=1$ or the control treatment $s=0$ with the randomization ratio $\phi$ and $1-\phi$, respectively.
	\item[1.2] During the $R^{\dagger}$th interim analysis with $R^{\dagger}\in (1, \cdots, R-1)$,  calculate ${\bf Z_{ R^{\dagger}}^{(o)}}$ and ${\rm PP}_{E, R^{\dagger}}$. 
	\item[1.3] If ${\rm PP}_{E, R^{\dagger}}<\lambda_E (\frac{N_{R^{\dagger}}}{N_R})^{\gamma_E}$, stop the trial and claim failure of the treatment $s=1$. If ${\rm PP}_{E, R^{\dagger}}>1-(1-\lambda_E) (\frac{N_{R^{\dagger}}}{N_R})^{\gamma_E}$, stop the trial and claim effectiveness of the treatment $s=1$. Otherwise, continue enrolling and randomizing patients until reaching the next interim analysis.
	\item[1.4] Repeat Steps~2 and~3 until the trial is either stopped early or reaches the final analysis. At the final analysis, update ${\bf Z_{R}^{(o)}}$ and $\mathrm{PP}_{E, R}$. Declare the treatment $s=1$ effective if $\mathrm{PP}_{E, R} > \lambda_E$; otherwise, deem the treatment ineffective.
	
\end{itemize}	

$\lambda_E (\frac{N_{R^{\dagger}}}{N_R})^{\gamma_E}$ and $1-(1-\lambda_E) (\frac{N_{R^{\dagger}}}{N_R})^{\gamma_E}$ are early-stopping functions for futility and superiority, respectively. $\lambda_E \in [0,1]$ and $\gamma_E \in [0,1]$ are design parameters, and their values can be calibrated through simulation studies to maximize power while controlling the type I error rate at a pre-specified nominal level $\alpha$. In particular, the calibration process can be summarized as follows:

\begin{itemize}
	\item[2.1] Specify $\alpha$, $\phi$, and the value of $p_{T}^{(n)}$ under the null hypothesis. 
	\item[2.2] By Theorem~1, generate $L$ i.i.d.\ samples of the vector $\mathbf{Z}_{R}$ (e.g., $L = 5{,}000$) from its asymptotic distribution under the null hypothesis with $\theta = 0$ and $p_{T} = p_{T}^{(n)}$. Denote these samples by $\mathbf{Z^{(n)}_{R,l}}$ for $l = 1, \ldots, L$. Then, by Theorem 2 and formula (5), for each $l$ samples calculate the corresponding vector of ${\rm PP}$ based on $\mathbf{Z^{(n)}_{R,l}}$ during all the interim and final analyses. 
	\item[2.3] Based on all the ${\rm PP}$s, estimate the probability of declaring treatment $s=1$ effective (POE) denoted by ${\rm POE}^{(n)}( \lambda_E, \gamma_E   )$.
	\item[2.4] Specify the values of $\theta^{(a)}$ and $p_{T}^{(a)}$ under the alternative hypothesis. Sampling $\mathbf{Z^{(a)}_{R,l}}$, calculate ${\rm PP}$s and estimate ${\rm POE}^{(a)}( \lambda_E, \gamma_E   )$ along the same line.
	\item[2.5] Conduct grid search and find the set of $(\lambda_E, \gamma_E)$ that achieve desirable type I error control with ${\rm POE}^{(n)}( \lambda_E, \gamma_E   )\leq \alpha$.
	\item[2.6] Among the set of $(\lambda_E, \gamma_E)$ identified in Step 5, select the optimal one that yields the maximum power of ${\rm POE}^{(a)}( \lambda_E, \gamma_E   )$. 
\end{itemize}	

We then extend the BMW design to incorporate toxicity outcomes. In general, toxicity outcomes are clinical different with the efficacy outcomes and cannot be directly integrated into the WR–based analysis. Therefore, it is more appropriate to evaluate the toxicity outcome  separately. In RCT that assess both efficacy and toxicity, a treatment is often preferred if it demonstrates superior efficacy while incurring only a clinically acceptable increase in toxicity compared with the control. Accordingly, we adopt a non-inferiority testing framework for toxicity.

We first outline the procedure for testing toxicity and then illustrate how to combine the efficacy and toxicity tests using a graphical testing procedure. Let $q_{T, s}$ denote the toxicity rate in arm $s$. Let $\delta>0$ denote the non-inferiority margin, which represents the maximum clinically acceptable increase in toxicity associated with the treatment compared to the control. The hypotheses for the toxicity are formulated as:

\begin{eqnarray}
	{\rm H}_{0,T}: q_{T,1}-q_{T,0}\geq \delta \ \ \  \leftrightarrow \ \ \ {\rm H}_{1,T}: q_{T,1}-q_{T,0}<\delta. 
\end{eqnarray}	

%
We assign independent noninformative Beta prior distributions $\mathrm{Beta}(1,1)$ to $q_{T,1}$ and $q_{T,0}$ separately. Let ${\cal D}_{T, R^{\dagger}}$ denote the accumulated toxicity data at the $R^{\dagger}$th interim analysis. The posterior probability (PP) for testing the non-inferiority of toxicity is then given by
\[
\mathrm{PP}_{T, R^{\dagger}} = \Pr(q_{T,1}-q_{T,0} < \delta \mid {\cal D}_{T, R^{\dagger}}).
\]

The same types of early-stopping boundaries, $\lambda_T \left(N_{R^{\dagger}}/N_R\right)^{\gamma_T}$ and $1 - (1-\lambda_T)\left(N_{R^{\dagger}}/N_R\right)^{\gamma_T}$, are applied at each interim and the final analysis for toxicity monitoring. The design parameters $\lambda_T$ and $\gamma_T$ are optimized following the same strategy used for $\lambda_E$ and $\gamma_E$. In particular, let $q_{T,0}^{(n)}$ denote the targeted value of $q_{T,0}$ under $H_{0, T}$, and let $q_{T,1}^{(a)}$ denote the targeted value of $q_{T, 1}$ under $H_{1,T}$, $\lambda_T$ and $\gamma_T$ are optimized to control  the type I error at level $\alpha$ given $q_{T,0}=q_{T,0}^{(n)}$ and $q_{T,1}=q_{T,0}^{(n)}+\delta$ while maximizing the power given $q_{T,0}=q_{T,0}^{(n)}$ and $q_{T,1}=q_{T,1}^{(a)}$. However, the numerical optimization for the design parameters for toxicity are conducted based on the observed data rather than from the test statistic.

We employ a graphical testing framework to test for both efficacy and toxicity outcomes. As illustrated in Figure 1, the primary objective is to test efficacy, and the test of toxicity is performed only after effectiveness of efficacy has been established. Once efficacy effectiveness is claimed, the entire nominal type I error level $\alpha$ is fully recycled to the subsequent toxicity test. Specifically, this design evaluates both efficacy and toxicity while controlling the family-wise error rate at level $\alpha$, and can be summarized as follows:

\begin{itemize}
	\item[3.1] Specify $\alpha$, $\delta$, $\phi$, $\theta^{(a)}$, $p_T^{(n)}$, $p_T^{(a)}$, $q_{T,0}^{(n)}$ and $q_{T,1}^{(a)}$, determine the optimal values of $(\lambda_E, \gamma_E)$ for efficacy and $(\lambda_T, \gamma_T)$ for toxicity, respectively. 
	\item[3.2] During the $R^{\dagger}$th interim analysis, calculate ${\rm PP}_{E, R^{\dagger}}$. If ${\rm PP}_{E, R^{\dagger}}<\lambda_E (\frac{N_{R^{\dagger}}}{N_R})^{\gamma_E}$, stop the trial and claim failure of the treatment $s=1$ due to the ineffectiveness. If ${\rm PP}_{E, R^{\dagger}}>1-(1-\lambda_E) (\frac{N_{R^{\dagger}}}{N_R})^{\gamma_E}$, go to steps 3.3 to 3.4. Otherwise, if the maximum sample size has not been reached, randomize more patients till the next interim analysis and repeat step 3.2; if the maximum sample size has been reached, go to step 3.5. 
	\item [3.3] Calculate ${\rm PP}_{T, R^{\dagger}}$. If ${\rm PP}_{T, R^{\dagger}}<\lambda_T (\frac{N_{R^{\dagger}}}{N_R})^{\gamma_T}$, stop the trial and claim failure of the treatment $s=1$ due to toxicity. If ${\rm PP}_{T, R^{\dagger}}>1-(1-\lambda_T) (\frac{N_{R^{\dagger}}}{N_R})^{\gamma_T}$, stop the trial and claim success of the treatment $s=1$. Otherwise, continue enrolling and randomizing patients until reaching the next interim analysis.
	\item[3.4] Repeat Step 3.3 until the trial is either stopped early or reaches the final analysis. At the final analysis, update $\mathrm{PP}_{T, R}$. Declare the success of the treatment $s=1$ if $\mathrm{PP}_{T, R} > \lambda_T$; otherwise, declare the failure of the treatment $s=1$ due to toxicity.
	\item[3.5] Update both $\mathrm{PP}_{E, R}$ and $\mathrm{PP}_{T, R}$. If ${\rm PP}_{E, R}>\lambda_E$ and ${\rm PP}_{T, R}>\lambda_T$, declare the success of the treatment $s=1$. If ${\rm PP}_{E, R}>\lambda_E$ and ${\rm PP}_{T, R}\leq \lambda_T$, declare the failure of the treatment $s=1$ due to toxicity. Otherwise, declare the failure of the treatment $s=1$ due to ineffectiveness.  
\end{itemize}

%
%

\section{Simulation}

In this section, we describe the simulation studies conducted to evaluate the operating characteristics (OCs) of the proposed BMW design. Two binary efficacy endpoints were considered: objective response (OR), denoted by $X_{E,1}=1$ (and $0$ otherwise), and event-free survival at 3 months (EFS3), denoted by $X_{E,2}=1$ (and $0$ otherwise). Let $q_{E,s}=(q_{E,1,s}, q_{E,2,s})$ denote the vector of the response probabilities for the efficacy endpoints $X_E=(X_{E,1}, X_{E,2})$ in treatment arm $s$. To generate correlated efficacy outcomes with marginal probabilities $q_{E,s}$, we introduced latent variables $W_E=(W_{E,1}, W_{E,2})$ following a bivariate normal distribution with mean vector $(0,0)$, variances 1, and correlation 0.25. Let $\mathrm{I}(\cdot)$ denote the indicator function. The observed binary efficacy outcomes were then defined as $X_{E} =\Big( \mathrm{I}\big(W_{E,1} \ge \kappa_{E,1,s}\big), \mathrm{I}\big(W_{E,2} \ge \kappa_{E,2,s}\big)\Big)$, where the cutoffs $\kappa_{E,s}=(\kappa_{E,1,s}, \kappa_{E,2,s})$ were selected to yield the desired marginal response probabilities $q_{E,s}$ under each simulation scenario. Toxicity outcomes were generated using the same latent-variable framework from latent variables $W_{ET}=(W_{E,1}, W_T)$ that followed a  similar bivariate normal distribution with correlation 0.2, thereby capturing the dependence between efficacy and toxicity outcomes.

The maximum sample size for the simulated trial was 160 patients, equally randomized to the new treatment and control arms. Two interim analyses were conducted after 80 and 120 patients had been treated, respectively. For the non-inferiority test of toxicity, the margin value was specified at $\delta = 0.1$. A nominal significance level of $\alpha = 0.1$ was used to control the type I error rate for the efficacy-only setting and FWER for the both efficacy and toxicity setting. For each simulation scenario, results were averaged based on 10,000 simulated trials.

We first consider the efficacy-only setting. Theorem 1 plays a central role in the development of the BMW design, as it enables direct derivation of the posterior probability from the WR statistic without requiring specification of the underlying data-generating mechanism. To empirically assess Theorem 1, we compared the BMW design with a raw-data-driven variant, denoted as the ${\rm BMW}_b$ design. Specifically, the ${\rm BMW}_b$ design assumes that the underlying data-generating mechanism is known and computes $\mathbf{Z}_R$ directly from simulated vectors of binary outcomes, rather than sampling from the asymptotic distribution in Steps 2.2 and 2.4 of the design parameter optimization procedure. Consequently, ${\rm BMW}_b$ serves as a benchmark, as it is used to evaluate the accuracy of the asymptotic distribution characterized in Theorem 1.
   
Table~1 summarizes the type~I error rate, power, and sample size under the BMW and ${\rm BMW}_b$ designs. Let $q^{(true)}_{E,0}$ denote the vector of true efficacy response rates for the control arm, and let $q^{(true,n)}_{E,1}$ and $q^{(true,a)}_{E,1}$ denote the corresponding vectors for the new treatment arm under the null and alternative hypotheses, respectively. Across all simulation scenarios, the pairs $(q^{(true)}_{E,0}, q^{(true,n)}_{E,1})$ and $(q^{(true)}_{E,0}, q^{(true,a)}_{E,1})$ yield the logarithm of WR of $\theta=0$ and $\theta=0.5$, respectively. Overall, the BMW and ${\rm BMW}_b$ designs exhibit highly similar operating characteristics. Under the null hypothesis, both designs maintain type~I error rates close to the nominal level. For example, in Scenario~1.1, the type~I error rates are 10.0\% for BMW and 9.9\% for ${\rm BMW}_b$, with expected sample sizes of 106.8 and 105.5, respectively. Under the alternative hypothesis, the two designs also demonstrate nearly identical power and sample size performance, as illustrated in Scenario~2.1, where the power is 80.0\% for BMW and 80.6\% for ${\rm BMW}_b$. These results support the validity and accuracy of the asymptotic distribution of $\mathbf{Z}_R$ described in Theorem~1.

We further verify the optimal design parameters $\lambda$ and $\gamma$ for the BMW design using Scenario~1.1 in Table~1 as an illustrative example. Under the settings of Scenario~1.1, the optimal parameters were identified as $\lambda_E = 0.92$ and $\gamma_E = 0.90$. We then conducted a grid search over all possible combinations of $\lambda_E$ and $\gamma_E$ and depicted the corresponding empirical type~I error and power surfaces under the ${\rm BMW}_b$ design. The results are shown in Figure~2, where the blue regions indicate combinations for which the type~I error is controlled at the 0.1 level. As demonstrated in Figure~2, the selected values of $\lambda_E$ and $\gamma_E$ maintain type~I error control at the nominal level while achieving maximal power, which confirms the optimal design property of the BMW design.
        
Table 2 reports the operating characteristics (OCs) of the BMW designs, including the rejection rate and the sample size. The rejection rate corresponds to the type I error under the null hypothesis (indicated by $^{*}$) and to power under the alternative hypothesis (indicated by $^{\dagger}$). We considered two versions of the BMW design: the original design and a variant that allows stopping for futility only, denoted by ${\rm BMW}_f$. For comparison, we also included a conventional method with no interim analyses, which uses the U-statistics test at the final analysis. As shown in Table 2, all three methods adequately control the type I error at the nominal level across all null scenarios, with rejection rates generally ranging between 8\% and 11\%. Under the alternative hypotheses, the three methods achieve broadly comparable power, typically between 75\% and 85\%, although the ${\rm BMW}_f$ design yields slightly higher power than the original BMW design in most scenarios. For example, in Scenario 3.5, the power of ${\rm BMW}_f$ is 85.4\%, compared with 80.2\% for BMW and 80.5\% for the U-statistic method.

With respect to sample size, the BMW design consistently requires the fewest patients among all methods. Under the null hypotheses, the sample sizes of ${\rm BMW}_f$ and BMW are generally comparable, both substantially smaller than that of the U-statistics method. For instance, in Scenario 1.1, the average sample sizes are 112.0 for ${\rm BMW}_f$ and 105.5 for BMW, compared with the fixed sample size of 160 for the U-statistics method. Under the alternative hypotheses, the difference of the sample sizes become substantial due to the BMW design's ability to stop early for efficacy. For example, in Scenario 2.4, BMW uses an average of 106.4 patients, compared with 153.2 for ${\rm BMW}_f$ and 160 for the U-statistic method. As expected, the U-statistic design always uses the maximum sample size because it does not allow for early stopping.

In summary, while all methods demonstrate similar operating characteristics in terms of type I error control and power, the BMW design offers substantial gains in sample size efficiency, particularly under the alternative hypothesis. Therefore, when interim monitoring is feasible, the BMW design is recommended due to its ability to markedly reduce sample size without sacrificing statistical performance.

We then consider the setting in which both efficacy and toxicity outcomes are evaluated. We compare the proposed BMW design with a conventional approach that does not incorporate interim analyses and conducts a single final analysis at the end of the trial. The conventional method uses the U-statistics test for efficacy and Fisher’s exact test for toxicity, and applies the same graphical testing procedure as that used in the BMW design. We report the FWER, the probability of correct selection (PCS), and the  sample size. The PCS is defined as the probability of making a correct go/no-go recommendation for the new treatment. In different with the traditional definition of family-wise power that is defined as the probability of rejecting at least one false null hypothesis, the PCS is more clinically relevant. For example, if the new treatment is ineffective but safe, and a no-go decision is correctly made based on the efficacy analysis, this result contributes to the PCS but not to the family-wise power, since no false null hypothesis (for toxicity) is rejected during the testing procedure. Therefore, we use the PCS as the evaluation criterion.  

Table 3 summarizes the operating characteristics of the BMW design and the conventional (Conv.) method across a wide range of efficacy–toxicity scenarios. Overall, both methods are able to well control the FWER around the nominal level typically below 10\%, which ensures the validity of the underlying graphical testing procedure. In terms of PCS, the two methods yield largely comparable performance. In scenarios where the treatment is both effective and safe (e.g., Scenarios 1.1), the Conv. method shows slightly higher PCS than the BMW design (e.g., 42.9\% vs. 37.4\% in Scenario 1.1), although the differences are at most modest. In contrast, when at least one endpoint fails, the PCS of the BMW design is very similar to, and sometimes even slightly higher than, that of the Conv. method (e.g., 97.0\% vs. 95.6\% in Scenario 1.3). Moreover, the BMW design consistently requires substantially fewer patients than the Conv. method, which fixes the sample size at 160. For example, when the treatment is ineffective but safe (Scenario 1.3),  the BMW design uses about 107 patients on average compared to 160 for the Conv. method, and similar reductions are observed throughout other ineffective scenarios (often saving more than 30–50 patients). Given its reliable FWER control, comparable power, and substantial reduction in sample size especially when the treatment is ineffective, the BMW design is recommended over the conventional approach.    

We conduct additional simulation studies to assess the sensitivity of the BMW design to several of its design parameters, with results reported in the online supplementary material. Table S1 summarizes the OCs when the probabilities of tie, $p_T^{(n)}$ and $p_T^{(a)}$, are mis-specified. These results demonstrate that the BMW design is robust to mis-specification of the tie probabilities. Table S2 examines the impact of varying the number of interim analyses, showing that fewer interim analyses can improve power but at the cost of a larger expected sample size. Finally, Table S3 considers a superiority test for toxicity in place of the non-inferiority test used in Table 3. The conclusions remain unchanged. That is, the BMW design requires substantially fewer patients while achieving comparable performance in other OCs compared with the conventional method.

\section{Software Development}

To facilitate the implementation of the BMW design in clinical practice, we develop an easy-to-use R Shiny app, which is freely available at https://iusccc.shinyapps.io/SmartDesign-Indiana/. Figure 3 shows the graphical user interface of the app. { \textcolor{red}{To be continued...} }

\section{Discussion}

In this paper, we propose the BMW design, a novel phase II Bayesian model-assisted design tailored to the WR statistic. The BMW design is highly flexible, accommodating arbitrary interim analyses with the option of early stopping for either superiority or futility. Unlike the BOP2 class of designs, which rely on simulated patient-level response data for calibration, the BMW design uses the joint asymptotic distribution of the WR statistic across multiple analysis times to optimize its design parameters. As a result, the BMW design is both efficient and robust, without requiring assumptions about the underlying data-generating mechanism. We further extend the BMW design to jointly evaluate efficacy and toxicity endpoints through a graphical testing procedure. The BMW design is fully transparent, with all decision rules pre-tabulated and readily incorporated into the trial protocol. Software for trial implementation and simulation studies is provided to facilitate practical use.

To implement the BMW design, the tie probabilities under the null and alternative hypotheses, denoted by \( p_T^{(n)} \) and \( p_T^{(a)} \), must be specified in advance. This requirement may pose challenges when prior information on these quantities is unavailable. A practical solution is to estimate them using in-trial data at the first interim analysis. Specifically, \( p_T^{(a)} \) can be estimated directly from the observed data. To estimate \( p_T^{(n)} \), a permutation procedure is needed first in which the treatment and control labels are randomly permuted, thereby generating data consistent with the null hypothesis.

The BMW design can be extended in several directions. First, when the target value of the logarithm of the WR, $\theta$, under the alternative hypothesis cannot be pre-specified, a conditional power framework may be employed to adapt the sample size at interim analyses \citep{meh11}. Second, the design can be extended to incorporate biomarker-defined subgroups, thereby accommodating population heterogeneity \citep{lip17}. Finally, it is of interest to generalize the BMW design to seamless phase II/III clinical trials with correlated endpoints \citep{sta11}.

\newpage

\begin{figure}[ht]
    \centering
    \includegraphics[width=\textwidth]{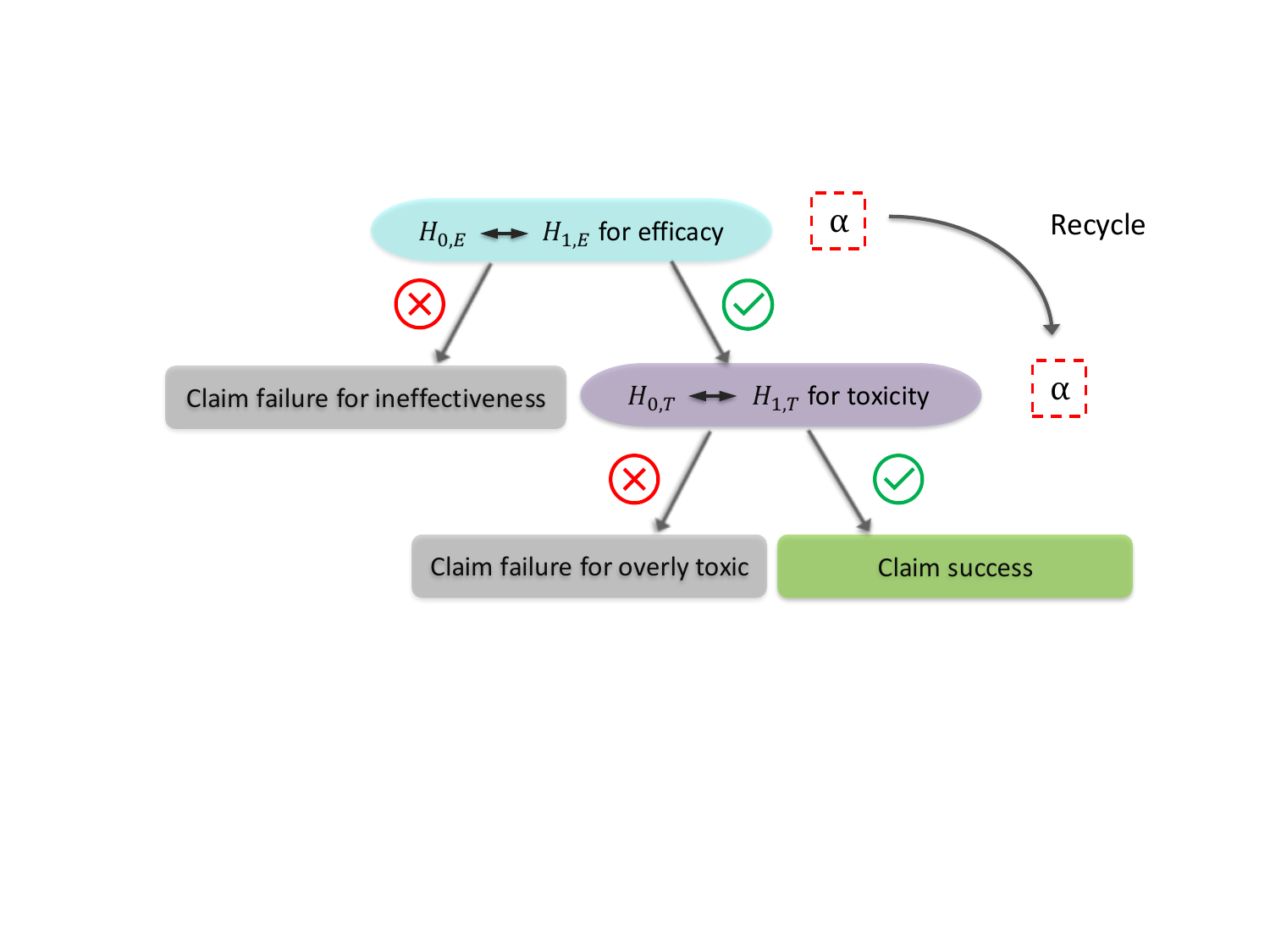}
    \caption{Schematic for the BMW design using the graphical testing procedure.}
    \label{fig1}
\end{figure}

\clearpage

\newpage 

\begin{figure}[ht]  
    \centering
    \includegraphics[width=\textwidth]{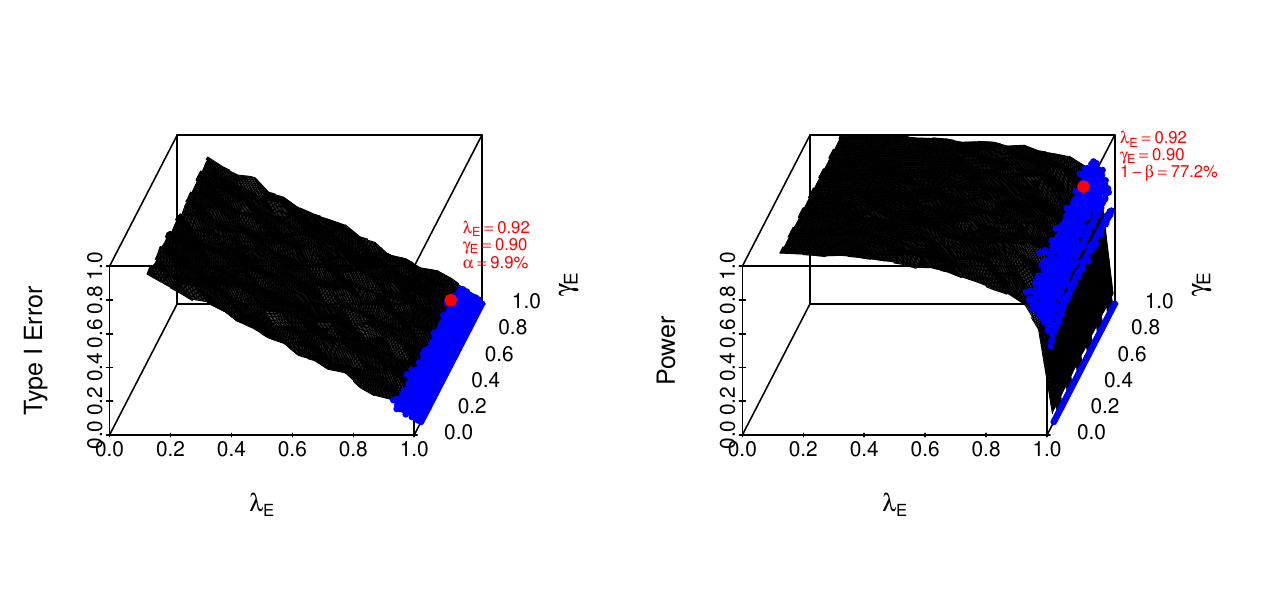}
    \caption{The type I error and power surfaces of the ${\rm BMV_b}$ design with different values of $\lambda$ and $\gamma$. The blue regions indicate that the type I error is controlled at 0.1. The red circles indicate the type I error and power with the optimal design parameters used by the BMW design.}
    \label{fig2}
\end{figure}

\begin{table}[htbp]

\caption{Type I error, power and sample size under the ${\rm BMW}$ and ${\rm BMW_b}$ designs evaluating efficacy outcomes only. $q^{(true)}_{E,0}$ denotes the vector of true efficacy response rates for the control arm. $q^{(true,n)}_{E,1}$ and $q^{(true,a)}_{E,1}$ denote the vectors of true efficacy response rates for the experimental treatment arm under the null and alternative hypotheses, respectively.}
\centering
\label{tab:oc_BMW_FS}
\setlength{\tabcolsep}{2.0 pt}
\renewcommand{\arraystretch}{1.12}
\begin{threeparttable}
\begin{adjustbox}{center, max width=\textwidth}
\begin{tabular}{c c c cc cc c cc cc}
\toprule
 & &  \multicolumn{5}{c}{\textbf{Null Hypothesis}} &  \multicolumn{5}{c}{\textbf{Alternative Hypothesis}} \\
\cmidrule(lr){3-7}\cmidrule(lr){8-12}
\textbf{Scenario} &$q^{(true)}_{E,0}$ & $q^{(true,n)}_{E,1}$ &
\multicolumn{2}{c}{\textbf{Type I Error (\%)}} &
\multicolumn{2}{c}{\textbf{Sample Size}} &
$q^{(true,a)}_{E,1}$  &
\multicolumn{2}{c}{\textbf{Power (\%)}} &
\multicolumn{2}{c}{\textbf{Sample Size}} \\
\cmidrule(lr){4-5}\cmidrule(lr){6-7}
\cmidrule(lr){9-10}\cmidrule(lr){11-12}
&  &  &
${\rm BMW}$ &${\rm BMW_b}$ &
${\rm BMW}$ & ${\rm BMW_b}$ &
&
${\rm BMW}$ & ${\rm BMW_b}$&
${\rm BMW}$ & ${\rm BMW_b}$ \\
\midrule
1.1 & \multirow{3}{*}{(0.40, 0.30)} & (0.40, 0.30) & 10.0 & 9.9 & 106.8 & 105.5 & (0.40, 0.66) & 79.8 & 77.2 & 109.0 & 108.7 \\
1.2 &  & (0.45, 0.21) & 9.8 & 9.3 & 107.2 & 105.8 & (0.45, 0.57) & 78.4 & 75.7 & 111.4 & 111.3 \\
1.3 &  & (0.50, 0.11) & 9.9 & 10.4 & 104.8 & 105.2 & (0.50, 0.48) & 77.4 & 75.6 & 109.8 & 109.9 \\
\midrule
2.1 & \multirow{3}{*}{(0.45, 0.35)} & (0.45, 0.35) & 8.6 & 8.5 & 103.0 & 101.8 & (0.45, 0.73) & 80.0 & 80.6 & 108.7 & 106.4 \\
2.2 &  & (0.50, 0.26) & 9.5 & 10.0 & 107.9 & 106.6 & (0.50, 0.63) & 80.0 & 82.6 & 110.1 & 108.9 \\
2.3 &  & (0.55, 0.16) & 9.8 & 8.9 & 106.5 & 106.2 & (0.55, 0.54) & 79.8 & 77.8 & 111.4 & 108.7 \\
\midrule
3.1 & \multirow{3}{*}{(0.50, 0.40)} & (0.50, 0.40) & 8.8 & 9.8 & 109.0 & 109.2 & (0.50, 0.78) & 80.2 & 82.1 & 111.3 & 110.1 \\
3.2 &  & (0.55, 0.31) & 9.9 & 10.2 & 103.9 & 103.1 & (0.55, 0.68) & 80.3 & 80.2 & 109.3 & 109.2 \\
3.3 &  & (0.60, 0.22) & 9.5 & 10.4 & 107.8 & 106.8 & (0.60, 0.58) & 78.3 & 80.3 & 111.0 & 111.5 \\
\midrule
4.1 & \multirow{3}{*}{(0.55, 0.45)} & (0.55, 0.45) & 9.6 & 10.1 & 106.6 & 107.4 & (0.55, 0.82) & 78.6 & 81.1 & 110.3 & 107.2 \\
4.2 &  & (0.60, 0.36) & 9.0 & 10.1 & 109.3 & 106.9 & (0.60, 0.72) & 78.8 & 80.1 & 111.8 & 109.2 \\
4.3 &  & (0.65, 0.27) & 9.6 & 9.9 & 103.1 & 103.6 & (0.65, 0.62) & 77.3 & 80.1 & 109.7 & 107.9 \\
\midrule
5.1 & \multirow{3}{*}{(0.60, 0.50)} & (0.60, 0.50) & 9.6 & 8.9 & 105.8 & 104.7 & (0.60, 0.85) & 77.5 & 78.6 & 112.0 & 110.5 \\
5.2 &  & (0.65, 0.41) & 9.0 & 10.2 & 107.3 & 104.7 & (0.65, 0.75) & 76.4 & 77.6 & 111.6 & 111.1 \\
5.3 &  & (0.70, 0.33) & 8.8 & 10.3 & 101.4 & 102.0 & (0.70, 0.66) & 75.4 & 77.0 & 109.2 & 108.3 \\
\bottomrule
\end{tabular}
\end{adjustbox}
\end{threeparttable}
\end{table}


\begin{table}[htbp]
	\centering
	\caption{Reject rate and sample size under the ${\rm BMW_f}$,  BMW designs and the conventional U-statistics method. $q^{(true)}_{E,0}$ and $q^{(true)}_{E,1}$ denotes the vector of true efficacy response rates for the control and new treatment arms, respectively.}
	\label{tab:oc_BMW_U}
	\setlength{\tabcolsep}{2.0 pt}
	\renewcommand{\arraystretch}{1.12}
	\begin{threeparttable}
		\begin{adjustbox}{center, max width=\textwidth}
			\begin{tabular}{c c c c c c c c c}
				\toprule
				\textbf{Scenario} &$q^{(true)}_{E,0}$& $q^{(true)}_{E,1}$ &
				\multicolumn{3}{c}{\textbf{Reject Rate (\%)}} &
				\multicolumn{3}{c}{\textbf{Sample Size}} \\
				\cmidrule(lr){4-6}\cmidrule(lr){7-9}
				& & & ${\rm BMW_f}$ & BMW & U
				& ${\rm BMW_f}$& BMW & U \\
				\midrule
				1.1$^{*}$ & \multirow{6}{*}{(0.40, 0.30)} & (0.40, 0.30)&  9.7 &  9.9 &  9.3 & 112.0 & 105.5 & 160.0 \\
				1.2$^{*}$ &                                   & (0.45, 0.21)& 10.1 &  9.3 &  9.1 & 114.4 & 105.8 & 160.0 \\
				1.3$^{*}$ &                                   & (0.50, 0.11) & 10.1 & 10.4 &  8.8 & 111.0 & 105.2 & 160.0 \\
				1.4$^{\dagger}$ &                              & (0.40, 0.66) &  78.3  & 77.2 & 79.1 & 153.4 & 108.7 & 160.0 \\
				1.5$^{\dagger}$ &                              & (0.45, 0.57)& 79.8 & 75.7 & 80.3& 154.2 & 111.3 & 160.0 \\
				1.6$^{\dagger}$ &                              & (0.50, 0.48)& 78.3 & 75.6 & 78.4 & 151.4 & 109.9 & 160.0 \\
				\midrule
				2.1$^{*}$ & \multirow{6}{*}{(0.45, 0.35)} & (0.45, 0.35)&  8.9 &  8.5 &  9.0 & 107.4 & 101.8 & 160.0 \\
				2.2$^{*}$ &                                   & (0.50, 0.26)&  8.9 & 10.0 &  9.3 & 109.5 & 106.6 & 160.0 \\
				2.3$^{*}$ &                                   & (0.55, 0.16) & 10.6 &  8.9 &  9.3 & 113.9 & 106.2 & 160.0 \\
				2.4$^{\dagger}$ &                              & (0.45, 0.73)& 82.3 & 80.6 & 81.4 & 153.2 & 106.4 & 160.0 \\
				2.5$^{\dagger}$ &                              & (0.50, 0.63)& 82.5 & 82.6 & 81.3 & 153.9 & 108.9 & 160.0 \\
				2.6$^{\dagger}$ &                              & (0.55, 0.54)& 82.5 & 77.8 & 79.5 & 154.3 & 108.7 & 160.0 \\
				\midrule
				3.1$^{*}$ & \multirow{6}{*}{(0.50, 0.40)} & (0.50, 0.40) & 10.5 &  9.8 &  9.3 & 110.1 & 109.2 & 160.0 \\
				3.2$^{*}$ &                                   & (0.55, 0.31) & 10.3 & 10.2 &  9.6 & 114.2 & 103.1 & 160.0 \\
				3.3$^{*}$ &                                   & (0.60, 0.22) & 10.7 & 10.4 &  9.0 & 112.2 & 106.8 & 160.0 \\
				3.4$^{\dagger}$ &                              & (0.50, 0.78) & 83.7 & 82.1 & 80.3 & 153.9 & 110.1 & 160.0 \\
				3.5$^{\dagger}$ &                              & (0.55, 0.68) & 85.4 & 80.2 & 80.5 & 155.2 & 109.2 & 160.0 \\
				3.6$^{\dagger}$ &                              & (0.60, 0.58) & 84.5 & 80.3 & 78.8 & 154.7 & 111.5 & 160.0 \\
				\midrule
				4.1$^{*}$ & \multirow{6}{*}{(0.55, 0.45)} & (0.55, 0.45) & 10.9 & 10.1 &  8.4 & 111.9 & 107.4 & 160.0 \\
				4.2$^{*}$ &                                   & (0.60, 0.36) &  9.4 & 10.1 &  8.9 & 113.1 & 106.9 & 160.0 \\
				4.3$^{*}$ &                                   & (0.65, 0.27) &  9.2 &  9.9 &  8.6 & 109.4 & 103.6 & 160.0 \\
				4.4$^{\dagger}$ &                              & (0.55, 0.82) & 82.8 & 81.1 & 78.5 & 153.9 & 107.2 & 160.0 \\
				4.5$^{\dagger}$ &                              & (0.60, 0.72) & 80.3 & 80.1 & 78.0 & 154.3 & 109.2 & 160.0 \\
				4.6$^{\dagger}$ &                              & (0.65, 0.62) & 80.7 & 80.1 & 78.3 & 153.0 & 107.9 & 160.0 \\
				\midrule
				5.1$^{*}$ & \multirow{6}{*}{(0.60, 0.50)} & (0.60, 0.50) &  9.4 &  8.9 &  8.5 & 107.6 & 104.7 & 160.0 \\
				5.2$^{*}$ &                                   & (0.65, 0.41) & 10.5 & 10.2 &  8.4 & 113.3 & 104.7 & 160.0 \\
				5.3$^{*}$ &                                   & (0.70, 0.33) & 10.8 & 10.3 &  8.2 & 107.4 & 102.0 & 160.0 \\
				5.4$^{\dagger}$ &                              & (0.60, 0.85) & 80.1 & 78.6 & 78.0 & 152.7 & 110.5 & 160.0 \\
				5.5$^{\dagger}$ &                              & (0.65, 0.75) & 80.8 & 77.6 & 76.6 & 154.2 & 111.1 & 160.0 \\
				5.6$^{\dagger}$ &                              & (0.70, 0.66) & 79.7 & 77.0 & 78.6 & 151.8 & 108.3 & 160.0 \\
				\bottomrule
			\end{tabular}
		\end{adjustbox}
		
		\begin{tablenotes}[flushleft]
			{\footnotesize
				\item $^{*}$: Type I error;
				\item $^{\dagger}$: Power.}
		\end{tablenotes}
		
	\end{threeparttable}
\end{table}

\begin{table}[htbp]
	\centering
	\caption{Family-wise error rate (FWER), probability of correction selection (PCS) and sample size under the BMW design and  the conventional method (Conv.). $q^{(true)}_{E,0}$ and $q^{(true)}_{E,1}$ denote the vectors of true efficacy response rates for the control and new treatment arms, respectively. $q^{(true)}_{T,0}$ and $q^{(true)}_{T,1}$ denote the counterparts for the toxicity.}
	\label{tab:compare_bmw_utox}
	
	\setlength{\tabcolsep}{2.0 pt}
	\renewcommand{\arraystretch}{0.7}
	\begin{small}
	\begin{threeparttable}
		\begin{adjustbox}{center, max width=\textwidth}
			\begin{tabular}{l cc cc cc cc cc}
				\toprule
				& \multicolumn{2}{c}{\textbf{Efficacy}}
				& \multicolumn{2}{c}{\textbf{Toxicity}}
				& \multicolumn{2}{c}{\textbf{FWER (\%)}}
				& \multicolumn{2}{c}{\textbf{PCS (\%)}}
				& \multicolumn{2}{c}{\textbf{Sample Size}} \\
				\cmidrule(lr){2-3}
				\cmidrule(lr){4-5}
				\cmidrule(lr){6-7}
				\cmidrule(lr){8-9}
				\cmidrule(lr){10-11}
				\textbf{Scenario}
				& $q^{(true)}_{E,0}$& $q^{(true)}_{E,1}$
				& $q^{(true)}_{T,0}$ & $q^{(true)}_{T,1}$
				& BMW & Conv.
				& BMW & Conv.
				& BMW & Conv.\\
				\midrule
				
				1.1$^{\dagger}$
				& \multirow{4}{*}{(0.40, 0.30)} & (0.40, 0.66)
				& \multirow{4}{*}{0.30} & 0.30
				& NA  & NA
				& 37.4 & 42.9
				& 129.5 & 160.0 \\
				1.2$^{\ddagger}$
				&                          & (0.40, 0.66)
				&                          & 0.40
				& 5.4  & 8.3
				& 94.6 & 91.7
				& 122.0 & 160.0 \\
				1.3$^{\mathsection}$
				&                          & (0.40, 0.30)
				&                          & 0.30
				& 7.2  & 9.5
				& 97.0 & 95.6
				& 107.4& 160.0 \\
				1.4$^{\mathparagraph}$
				&                          & (0.40, 0.30)
				&                          & 0.40
				& 9.8  & 9.9
				& 99.6 & 99.0
				& 107.6 & 160.0 \\
				\midrule
				
				2.1$^{\dagger}$
				& \multirow{4}{*}{(0.40, 0.30)} & (0.50, 0.48)
				& \multirow{4}{*}{0.30} & 0.30
				& NA  & NA
				& 35.4 & 40.8
				& 131.0 & 160.0 \\
				2.2$^{\ddagger}$
				&                          & (0.50, 0.48)
				&                          & 0.40
				& 6.4  & 7.2
				& 93.6 & 92.8
				& 123.9 & 160.0 \\
				2.3$^{\mathsection}$
				&                          & (0.50, 0.11)
				&                          & 0.30
				& 9.1  & 9.6
				& 96.5 & 95.4
				& 108.3 & 160.0 \\
				2.4$^{\mathparagraph}$
				&                          & (0.50, 0.11)
				&                          & 0.40
				& 9.9 & 9.5
				& 98.9 & 99.8
				& 106.8 & 160.0 \\
				\midrule
				
				3.1$^{\dagger}$
				& \multirow{4}{*}{(0.45, 0.35)} & (0.45, 0.73)
				& \multirow{4}{*}{0.35} & 0.35
				& NA  & NA
				& 38.1 & 42.6
				& 129.4 & 160.0 \\
				3.2$^{\ddagger}$
				&                          & (0.45, 0.73)
				&                          & 0.45
				& 6.9  & 8.6
				& 93.1 & 91.4
				& 121.6& 160.0 \\
				3.3$^{\mathsection}$
				&                          & (0.45, 0.35)
				&                          & 0.35
				& 9.6  & 9.7
				& 96.4 & 95.6
				& 106.5 & 160.0 \\
				3.4$^{\mathparagraph}$
				&                          & (0.45, 0.35)
				&                          & 0.45
				& 10.1  & 10.1
				& 99.2 & 99.3
				& 104.6 & 160.0 \\
				\midrule
				
				4.1$^{\dagger}$
				& \multirow{4}{*}{(0.45, 0.35)} & (0.55, 0.54)
				& \multirow{4}{*}{0.35} & 0.35
				& NA   & NA
				& 36.4 & 41.2
				& 131.1 & 160.0 \\
				4.2$^{\ddagger}$
				&                          & (0.55, 0.54)
				&                          & 0.45
				& 7.3  & 8.4
				& 92.7 & 91.6
				& 124.7 & 160.0 \\
				4.3$^{\mathsection}$
				&                          & (0.55, 0.16)
				&                          & 0.35
				& 9.6  & 9.7
				& 96.2 & 95.7
				& 109.5& 160.0 \\
				4.4$^{\mathparagraph}$
				&                          & (0.55, 0.16)
				&                          & 0.45
				& 9.2  & 10.3
				& 99.6 & 99.2
				& 109.0 & 160.0 \\
				\midrule
				
				5.1$^{\dagger}$
				& \multirow{4}{*}{(0.50, 0.40)} & (0.50, 0.80)
				& \multirow{4}{*}{0.35} & 0.35
				& NA   & NA
				& 37.6 & 41.7
				& 130.8 & 160.0 \\
				5.2$^{\ddagger}$
				&                          & (0.50, 0.80)
				&                          & 0.45
				& 7.9  & 7.6
				& 92.1 & 92.4
				& 124.4& 160.0 \\
				5.3$^{\mathsection}$
				&                          & (0.50, 0.40)
				&                          & 0.35
				& 9.1  & 8.9
				& 95.9 & 96.0
				& 109.2 & 160.0 \\
				5.4$^{\mathparagraph}$
				&                          & (0.50, 0.40)
				&                          & 0.45
				& 9.4  & 9.7
				& 99.8 & 99.3
				& 108.5 & 160.0 \\
				\midrule
				
				6.1$^{\dagger}$
				& \multirow{4}{*}{(0.50, 0.40)} & (0.60, 0.58)
				& \multirow{4}{*}{0.35} & 0.35
				& NA  & NA
				& 36.3 & 41.5
				& 129.7 & 160.0 \\
				6.2$^{\ddagger}$
				&                          & (0.60, 0.58)
				&                          & 0.45
				& 7.6  & 8.0
				& 92.4 & 92.0
				& 122.8 & 160.0 \\
				6.3$^{\mathsection}$
				&                          & (0.60, 0.22)
				&                          & 0.35
				& 8.2  & 9.9
				& 96.5 & 95.8
				& 105.7 & 160.0 \\
				6.4$^{\mathparagraph}$
				&                          & (0.60, 0.22)
				&                          & 0.45
				& 9.3 & 9.9
				& 99.8 & 99.3
				& 105.8 & 160.0 \\
				\midrule
				7.1$^{\dagger}$
				& \multirow{4}{*}{(0.55, 0.45)} & (0.55, 0.82)
				& \multirow{4}{*}{0.40} & 0.40
				& NA  & NA
				& 38.1 & 40.6
				& 132.3 & 160.0 \\
				7.2$^{\ddagger}$
				&                          & (0.55, 0.82)
				&                          & 0.50
				& 7.4  & 6.5
				& 92.6 & 93.5
				& 126.2 & 160.0 \\
				7.3$^{\mathsection}$
				&                          & (0.55, 0.45)
				&                          & 0.40
				& 6.1 & 9.7
				& 97.6  & 95.3
				& 110.5 & 160.0 \\
				7.4$^{\mathparagraph}$
				&                          & (0.55, 0.45)
				&                          & 0.50
				& 7.7  & 9.8
				& 99.2 & 99.5
				& 108.8 & 160.0 \\
				\midrule
				
				8.1$^{\dagger}$
				& \multirow{4}{*}{(0.55, 0.45)} & (0.65, 0.62)
				& \multirow{4}{*}{0.40} & 0.40
				& NA  & NA
				& 34.5 & 39.9
				& 130.7 & 160.0 \\
				8.2$^{\ddagger}$
				&                          & (0.65, 0.62)
				&                          & 0.50
				& 6.8  & 7.7
				& 93.2 & 92.3
				& 124.8 & 160.0 \\
				8.3$^{\mathsection}$
				&                          & (0.65, 0.27)
				&                          & 0.40
				& 7.4  & 8.9
				& 97.1 & 96.4
				& 108.4 & 160.0 \\
				8.4$^{\mathparagraph}$
				&                          & (0.65, 0.27)
				&                          & 0.50
				& 8.3 & 9.6
				& 99.6 & 99.6
				& 107.6 & 160.0 \\
				\midrule
				
				9.1$^{\dagger}$
				& \multirow{4}{*}{(0.60, 0.50)} & (0.60, 0.85)
				& \multirow{4}{*}{0.40} & 0.40
				& NA  & NA
				& 35.8  & 38.4
				& 130.3 & 160.0 \\
				9.2$^{\ddagger}$
				&                          & (0.60, 0.85)
				&                          & 0.50
				& 6.1  & 7.4
				& 93.9 & 92.6
				& 126.4 & 160.0 \\
				9.3$^{\mathsection}$
				&                          & (0.60, 0.50)
				&                          & 0.40
				& 8.4  & 9.0
				& 95.6 & 96.3
				& 110.6 & 160.0 \\
				9.4$^{\mathparagraph}$
				&                          & (0.60, 0.50)
				&                          & 0.50
				& 9.2  & 9.1
				& 99.4 & 99.2
				& 108.0 & 160.0 \\
				\midrule
				
				10.1$^{\dagger}$
				& \multirow{4}{*}{(0.60, 0.50)} & (0.70, 0.66)
				& \multirow{4}{*}{0.40} & 0.40
				& NA  & NA
				& 36.1 & 37.3
				& 131.1 & 160.0 \\
				10.2$^{\ddagger}$
				&                           & (0.70, 0.66)
				&                           & 0.50
				& 8.3  & 6.7
				& 91.7 & 93.3
				& 125.3 & 160.0 \\
				10.3$^{\mathsection}$
				&                           & (0.70, 0.33)
				&                           & 0.40
				& 9.4 & 8.1
				& 95.6 & 96.7
				& 110.5 & 160.0 \\
				10.4$^{\mathparagraph}$
				&                           & (0.70, 0.33)
				&                           & 0.50
				& 9.9  & 8.9
				& 99.3 & 99.0
				& 106.7 & 160.0 \\
				\bottomrule
			\end{tabular}
		\end{adjustbox}
		
		\begin{tablenotes}[flushleft]
			{\footnotesize
				\item $^{\dagger}:$ $(\text{Efficacy pass},\,\text{Toxicity pass})$.
				\item $^{\ddagger}:$ $(\text{Efficacy pass},\,\text{Toxicity fail})$.
				\item $^{\mathsection}:$ $(\text{Efficacy fail},\,\text{Toxicity pass})$.
				\item $^{\mathparagraph}:$ $(\text{Efficacy fail},\,\text{Toxicity fail})$.
			}
		\end{tablenotes}
		
	\end{threeparttable}
\end{small}
\end{table}


\end{document}


\doublespacing

	\begin{center}
	{\bf \Large Supplementary Material for ``BMW: Bayesian Model-Assisted Adaptive Phase II Clinical Trial Design for Win Ratio Statistic''}
	\medskip
\end{center}

	\vskip.3in
\noindent {\bf Proof of Theorem 1. }

At the $r$th analysis, define the standardized win ratio statistic
\[
z_r=\frac{\widehat{\theta}_r}{\sqrt{{\rm Var}(\widehat{\theta}_r)}}
=\widehat{\theta}_r\sqrt{{\rm I}_r},
\qquad r=1,\ldots,R^{\dagger},
\]
where $\widehat{\theta}_r=\log({\rm WR}_r)$ and
\[
{\rm I}_r=\frac{1}{{\rm Var}(\widehat{\theta}_r)}
=\frac{3\phi(1-\phi)(1-p_T)N_r}{4(1+p_T)}.
\]

Under the regularity conditions for the win ratio estimator,
\[
(\widehat{\theta}_r-\theta)\sqrt{{\rm I}_r}
\xrightarrow{d}\mathcal N(0,1),
\qquad r=1,\ldots,R,
\]
which implies
\[
z_r \xrightarrow{d}\mathcal N(\mu_r,1),
\qquad \mu_r=\theta\sqrt{{\rm I}_r}.
\]

\medskip
We next derive the covariance structure of $(z_{r_1},z_{r_2})$ for $r_1<r_2$.
Define the centered standardized statistic
\[
\tilde z_r=(\widehat{\theta}_r-\theta)\sqrt{{\rm I}_r}.
\]
The Fisher information admits the representation
\[
{\rm I}_r=\left(
\frac{\sigma_1^2}{\phi N_r}
+
\frac{\sigma_2^2}{(1-\phi)N_r}
\right)^{-1},
\]
where
\[
\sigma_1^2={\rm Var}\{H_2(X_{1i})\},\qquad
\sigma_2^2={\rm Var}\{H_1(X_{2j})\},
\]
with
\begin{align*}
H_1(X_{2j})
&=\frac{1}{\phi N_r}\sum_{i=1}^{\phi N_r}
\mathbb I(X_{1i}>X_{2j}),\\
H_2(X_{1i})
&=\frac{1}{(1-\phi)N_r}\sum_{j=1}^{(1-\phi)N_r}
\mathbb I(X_{2j}>X_{1i}).
\end{align*}

For $r_1<r_2$, the data used at analysis $r_1$ are a subset of those used at
analysis $r_2$. Therefore,
\begin{align*}
{\rm Cov}(\tilde z_{r_1},\tilde z_{r_2})
&=
\sqrt{{\rm I}_{r_1}{\rm I}_{r_2}}\,
{\rm Cov}\Bigg[
\frac{1}{(1-\phi)N_{r_1}}\sum_{l=1}^{(1-\phi)N_{r_1}} H_1(X_{2l})
-
\frac{1}{\phi N_{r_1}}\sum_{m=1}^{\phi N_{r_1}} H_2(X_{1m}),\\
&\hspace{3.5cm}
\frac{1}{(1-\phi)N_{r_2}}\sum_{l'=1}^{(1-\phi)N_{r_2}} H_1(X_{2l'})
-
\frac{1}{\phi N_{r_2}}\sum_{m'=1}^{\phi N_{r_2}} H_2(X_{1m'})
\Bigg].
\end{align*}

Using independence of non-overlapping observations and the nested structure,
\begin{align*}
{\rm Cov}(\tilde z_{r_1},\tilde z_{r_2})
&=
\sqrt{{\rm I}_{r_1}{\rm I}_{r_2}}
\left(
\frac{1}{(1-\phi)^2 N_{r_1}N_{r_2}}\cdot(1-\phi)N_{r_1}\sigma_2^2
+
\frac{1}{\phi^2 N_{r_1}N_{r_2}}\cdot\phi N_{r_1}\sigma_1^2
\right)\\
&=
\sqrt{{\rm I}_{r_1}{\rm I}_{r_2}}
\left(
\frac{\sigma_2^2}{(1-\phi)N_{r_2}}
+
\frac{\sigma_1^2}{\phi N_{r_2}}
\right)
=
\sqrt{\frac{{\rm I}_{r_1}}{{\rm I}_{r_2}}}.
\end{align*}

Since $z_r=\tilde z_r+\theta\sqrt{{\rm I}_r}$ differs from $\tilde z_r$ only by a
mean shift, the same covariance holds for $z_{r_1}$ and $z_{r_2}$. Moreover,
${\rm Var}(z_r)=1$, and thus
\[
\rho_{r_1,r_2}
={\rm Corr}(z_{r_1},z_{r_2})
=\sqrt{\frac{{\rm I}_{r_1}}{{\rm I}_{r_2}}},
\qquad r_1<r_2.
\]

\medskip
Let $\mathbf{Z_{R^{\dagger}}}=(z_1,\ldots,z_{R^{\dagger}})^{\rm T}$.
Combining the marginal asymptotic normality and the above correlation structure,
we conclude that
\[
\mathbf{Z_{R^{\dagger}}}
\sim {\cal N}({\bf M_{R^{\dagger}}},{\bf \Sigma_{R^{\dagger}}}),
\]
where ${\bf M_{R^{\dagger}}}=(\mu_1,\ldots,\mu_{R^{\dagger}})^{\rm T}$ with
$\mu_r=\theta\sqrt{{\rm I}_r}$ and ${\bf \Sigma_{R^{\dagger}}}$ has unit diagonal
elements and off-diagonal elements
$\Sigma_{r_1,r_2}=\rho_{r_1,r_2}$ for $r_1<r_2$.
This completes the proof.

	\vskip.3in
\noindent {\bf Proof of Theorem 2. }

Assume a normal prior for the true log win ratio parameter $\theta$:
\[
\theta \sim \mathcal{N}(\theta_0,\sigma_0^2)\]
with prior density function:
\[
\pi(\theta)=\frac{1}{\sqrt{2\pi\sigma_0^2}}
\exp\left\{-\frac{1}{2\sigma_0^2}(\theta-\theta_0)^2\right\}.
\]

At the current interim $R^{\dagger}$, given $\theta$, the observed standardized statistics
$\mathbf{Z}_{R^{\dagger}}^{(o)}=(Z_1^{(o)},\ldots,Z_{R^{\dagger}}^{(o)})^\top$
follow a multivariate normal distribution
\[
\mathbf{Z}_{R^{\dagger}}^{(o)} \mid \theta
\sim
\mathcal{N}_{R^{\dagger}}\!\left(\theta\,\mathbf{B}_{R^{\dagger}},\,\Sigma_{R^{\dagger}}\right),
\qquad
\mathbf{B}_{R^{\dagger}}=(\sqrt{I_1},\ldots,\sqrt{I_{R^{\dagger}}})^\top .
\]

The likelihood function is therefore
\[
\mathcal{L}\!\left(\theta;\mathbf{Z}_{R^{\dagger}}^{(o)}\right)
\propto
\exp\left\{-\frac{1}{2}
\left(\mathbf{Z}_{R^{\dagger}}^{(o)}-\theta\mathbf{B}_{R^{\dagger}}\right)^\top
\Sigma_{R^{\dagger}}^{-1}
\left(\mathbf{Z}_{R^{\dagger}}^{(o)}-\theta\mathbf{B}_{R^{\dagger}}\right)
\right\}.
\]

The posterior density is proportional to the product of prior and likelihood:
\begin{align*}
f\!\left(\theta \mid \mathbf{Z}_{R^{\dagger}}^{(o)}\right)
&\propto
\exp\left\{-\frac{1}{2\sigma_0^2}(\theta-\theta_0)^2\right\}
\exp\left\{-\frac{1}{2}
\left(\mathbf{Z}_{R^{\dagger}}^{(o)}-\theta\mathbf{B}_{R^{\dagger}}\right)^\top
\Sigma_{R^{\dagger}}^{-1}
\left(\mathbf{Z}_{R^{\dagger}}^{(o)}-\theta\mathbf{B}_{R^{\dagger}}\right)
\right\} \\
&=
\exp\left\{-\frac{1}{2}\left[
\frac{(\theta-\theta_0)^2}{\sigma_0^2}
+
\left(\mathbf{Z}_{R^{\dagger}}^{(o)}-\theta\mathbf{B}_{R^{\dagger}}\right)^\top
\Sigma_{R^{\dagger}}^{-1}
\left(\mathbf{Z}_{R^{\dagger}}^{(o)}-\theta\mathbf{B}_{R^{\dagger}}\right)
\right]\right\}.
\end{align*}

Expanding the quadratic forms gives
\begin{align*}
\frac{(\theta-\theta_0)^2}{\sigma_0^2}
&=
\frac{\theta^2}{\sigma_0^2}
-\frac{2\theta_0\theta}{\sigma_0^2}
+\text{const.},\\
\left(\mathbf{Z}_{R^{\dagger}}^{(o)}-\theta\mathbf{B}_{R^{\dagger}}\right)^\top
\Sigma_{R^{\dagger}}^{-1}
\left(\mathbf{Z}_{R^{\dagger}}^{(o)}-\theta\mathbf{B}_{R^{\dagger}}\right)
&=
\theta^2\,\mathbf{B}_{R^{\dagger}}^\top\Sigma_{R^{\dagger}}^{-1}\mathbf{B}_{R^{\dagger}}
-2\theta\,\mathbf{B}_{R^{\dagger}}^\top\Sigma_{R^{\dagger}}^{-1}\mathbf{Z}_{R^{\dagger}}^{(o)}
+\text{const.}
\end{align*}

Combining all terms,
\[
f\!\left(\theta \mid \mathbf{Z}_{R^{\dagger}}^{(o)}\right)
\propto
\exp\left\{-\frac{1}{2}\left[
A\theta^2-2B\theta+\text{const.}
\right]\right\},
\]
where
\[
A=\frac{1}{\sigma_0^2}
+\mathbf{B}_{R^{\dagger}}^\top\Sigma_{R^{\dagger}}^{-1}\mathbf{B}_{R^{\dagger}},
\qquad
B=\frac{\theta_0}{\sigma_0^2}
+\mathbf{B}_{R^{\dagger}}^\top\Sigma_{R^{\dagger}}^{-1}\mathbf{Z}_{R^{\dagger}}^{(o)}.
\]

Completing the square yields
\[
f\!\left(\theta \mid \mathbf{Z}_{R^{\dagger}}^{(o)}\right)
\propto
\exp\left\{-\frac{A}{2}\left(\theta-\frac{B}{A}\right)^2\right\},
\]
which implies
\[
\theta \mid \mathbf{Z}_{R^{\dagger}}^{(o)}
\sim
\mathcal{N}\!\left(\theta_{R^{\dagger}},\,\sigma_{R^{\dagger}}^2\right),
\]
with posterior parameters
\[
\sigma_{R^{+}}^2=
\left(\frac{1}{\sigma_0^2}
+\mathbf{B}_{R^{\dagger}}^\top\Sigma_{R^{\dagger}}^{-1}\mathbf{B}_{R^{\dagger}}\right)^{-1},
\qquad
\theta_{R^{\dagger}}=
\sigma_{R^{\dagger}}^2\left(
\frac{\theta_0}{\sigma_0^2}
+\mathbf{B}_{R^{\dagger}}^\top\Sigma_{R^{\dagger}}^{-1}\mathbf{Z}_{R^{\dagger}}^{(o)}
\right).
\]

\begin{table}[htbp]
	\centering
	\renewcommand{\thetable}{S1}
	\caption{Sensitivity analysis to the probabilities of tie. $p_{T}^{(n,used)}$ and $p_T^{(a,used)}$ are the probabilities of tie used to generate the BMW design. $p_{T}^{(n,used)}$ and $p_T^{(a,used)}$ are the underlying true probabilities of tie for the data.}
	\label{tab:oc_bmwfs_tie}
	
	\setlength{\tabcolsep}{2pt}
	\renewcommand{\arraystretch}{1.12}
	
	\begin{threeparttable}
		\begin{adjustbox}{center, max width=0.95\textwidth}
			\begin{tabular}{c c c c c c c c c}
				\toprule
				&  \multicolumn{4}{c}{\textbf{Probabilities of tie}} & \multicolumn{2}{c}{\textbf{Null Hypothesis}} & \multicolumn{2}{c}{\textbf{Alternative Hypothesis}} \\
				\cmidrule(lr){2-5}\cmidrule(lr){6-7}\cmidrule(lr){8-9}
				\textbf{Scenario} & \textbf{$p_{T}^{(n, used)}$} & \textbf{$p_{T}^{(a, used)}$} & \textbf{$p_{T}^{(n,true)}$} & \textbf{$p_{T}^{(a,true)}$} &
				\textbf{Type I error (\%)} & \textbf{Sample Size} &
				\textbf{Power (\%)} & \textbf{Sample Size} \\
				\midrule
				
				1.1 & \multirow{5}{*}{0.31} & \multirow{5}{*}{0.23} & 0.31 & 0.23 & 9.9 & 105.5 & 77.2 & 108.7 \\
				1.2 &                      &                      & 0.28 & 0.21 & 10.0 & 104.4 & 82.8 & 107.2 \\
				1.3       &                      &                      & 0.33 & 0.26 & 10.4 & 104.0 & 79.5 & 109.7 \\
				1.4       &                      &                      & 0.28 & 0.26 & 12.5 & 107.5 & 79.5 & 109.7 \\
				1.5    &                      &                      & 0.33 & 0.21 & 10.4 & 104.0 & 82.8 & 107.2 \\
				\midrule
				
				2.1 & \multirow{5}{*}{0.28} & \multirow{5}{*}{0.22} & 0.28 & 0.22 & 8.5 & 101.8 & 80.6 & 106.4 \\
				2.2 &                      &                      & 0.26 & 0.20 & 10.0 & 107.3 & 84.1 & 107.0 \\
				2.3       &                      &                      & 0.30 & 0.25 & 9.4 & 106.9 & 81.8 & 109.7 \\
				2.4       &                      &                      & 0.26 & 0.25 & 12.0 & 107.0 & 79.5 & 110.7 \\
				2.5    &                      &                      & 0.30 & 0.20 & 11.5 & 106.8 & 83.5 & 107.1 \\
				\midrule
				
				3.1& \multirow{5}{*}{0.27} & \multirow{5}{*}{0.23} & 0.27 & 0.23 & 9.8 & 109.2 & 82.1 & 110.1 \\
				3.2 &                      &                      & 0.24 & 0.20 & 10.0 & 107.2 & 83.3 & 106.5 \\
				3.3      &                      &                      & 0.28 & 0.25 & 10.0 & 105.8 & 80.6 & 109.8 \\
				3.4      &                      &                      & 0.24 & 0.25 & 12.0 & 108.1 & 78.0 & 110.2 \\
				3.5    &                      &                      & 0.28 & 0.20 & 11.8 & 106.5 & 82.0 & 107.7 \\
				\midrule
				
				4.1 & \multirow{5}{*}{0.26} & \multirow{5}{*}{0.24} & 0.26 & 0.24 & 10.1 & 107.4 & 81.1 & 107.2 \\
				4.2 &                      &                      & 0.23 & 0.22 & 9.9 & 105.9 & 84.3 & 110.5 \\
				4.3      &                      &                      & 0.27 & 0.27 & 10.0 & 107.3 & 80.6 & 109.6 \\
				4.4      &                      &                      & 0.23 & 0.27 & 10.7 & 108.2 & 77.9 & 110.1 \\
				4.5    &                      &                      & 0.27 & 0.22 & 10.2 & 107.9 & 81.2 & 108.7 \\
				\midrule
				
				5.1 & \multirow{5}{*}{0.27} & \multirow{5}{*}{0.27} & 0.27 & 0.27 & 8.9 & 104.7 & 78.6 & 110.5 \\
				5.2 &                      &                      & 0.24 & 0.24 & 10.0 & 108.5 & 79.5 & 111.0 \\
				5.3       &                      &                      & 0.29 & 0.30 & 12.1 & 107.4 & 79.1 & 110.9 \\
				5.4      &                      &                      & 0.24 & 0.30 & 10.6 & 106.7 & 77.0 & 111.5 \\
				5.5    &                      &                      & 0.29 & 0.24 & 10.8 & 108.2 & 79.1 & 110.7 \\
				\bottomrule
			\end{tabular}
		\end{adjustbox}

	\end{threeparttable}
	
\end{table}

\begin{table}[htbp]
	\centering
	\renewcommand{\thetable}{S2}
	\caption{Sensitivity analysis to different numbers of interim analysis. ${\rm BMW_1}$ conducts one interim analysis after 80 patients have been treated. ${\rm BMW_2}$ conducts two interim analyses after 80 and 120 patients have been treated. ${\rm BMW_3}$ conducts three interim analyses after 30, 80 and 120 patients have been treated.}
	\label{tab:oc_bmwfs_interims}
	
	\setlength{\tabcolsep}{2.0 pt}
	\renewcommand{\arraystretch}{1.12}
	\begin{threeparttable}
		\begin{adjustbox}{center, max width=\textwidth}
			\begin{tabular}{c c c ccc ccc}
				\toprule
				&  &  &
				\multicolumn{3}{c}{\textbf{Reject Rate (\%)}} &
				\multicolumn{3}{c}{\textbf{Sample Size}} \\
				\cmidrule(lr){4-6}\cmidrule(lr){7-9}
				\textbf{Scenario} & \textbf{$q_{E,0}^{(true)}$} & \textbf{$q_{E,1}^{(true)}$} &
				${\rm BMW_1}$ & ${\rm BMW_2}$ & ${\rm BMW_3}$ &
			${\rm BMW_1}$ & ${\rm BMW_2}$ & ${\rm BMW_3}$  \\
				\midrule
				
				S1.1$^{\star}$     & \multirow{2}{*}{(0.40, 0.30)} & (0.40, 0.30) & 9.9  & 9.9  & 11.1  & 119.4 & 105.5 & 88.4  \\
				S1.2$^{\dagger}$      &  & (0.40, 0.66) & 77.7 & 77.2 & 74.8 & 121.0 & 108.7 & 99.6 \\
				\midrule
				
				S2.1$^{\star}$     & \multirow{2}{*}{(0.45, 0.35)} & (0.45, 0.35) & 9.3  & 8.5  & 8.9  & 116.6 & 101.8 & 93.2  \\
				S2.2$^{\dagger}$      &  & (0.45, 0.73) & 85.3 & 80.6 & 80.7 & 115.7 & 106.4 & 98.4\\
				\midrule
				
				S3.1$^{\star}$     & \multirow{2}{*}{(0.50, 0.40)} & (0.50, 0.40) & 11.0  & 9.8  & 9.9  & 120.3 & 109.2 & 94.9  \\
				S3.2$^{\dagger}$      &  & (0.50, 0.78) & 82.7 & 82.1 & 79.8 & 120.0 & 110.1 & 100.8 \\
				\midrule
				
				S4.1$^{\star}$     & \multirow{2}{*}{(0.55, 0.45)} & (0.55, 0.45) & 8.9  & 10.1  & 8.6  & 120.5 & 107.4 & 97.2  \\
				S4.2$^{\dagger}$      &  & (0.55, 0.82) & 80.6 & 81.1 & 78.8 & 119.0 & 107.2 & 98.4 \\
				\midrule
				
				S5.1$^{\star}$     & \multirow{2}{*}{(0.60, 0.50)} & (0.60, 0.50) & 9.9  & 8.9  & 9.7  & 119.5 & 104.7 & 86.9  \\
				S5.2$^{\dagger}$      &  & (0.60, 0.85) & 79.1 & 78.6 & 75.2 & 121.0 & 110.5 & 98.7 \\
				\bottomrule
			\end{tabular}
		\end{adjustbox}
		
		\begin{tablenotes}[flushleft]
			\footnotesize
			\item $^{\star}$: Type I error.
			\item $^{\dagger}$: Power.
		\end{tablenotes}

	\end{threeparttable}
\end{table}

\begin{table}[htbp]
	\centering
	\renewcommand{\thetable}{S3} 
	\caption{Sensitivity analysis to superiority test for toxicity.}
	\label{tab:compare_bmw_utox_sup}
	
	\setlength{\tabcolsep}{2.0 pt}
	\renewcommand{\arraystretch}{0.7}
	
	\begin{threeparttable}
		\begin{adjustbox}{center, max width=\textwidth}
			\begin{tabular}{l cc cc cc cc cc}
				\toprule
				& \multicolumn{2}{c}{\textbf{Efficacy}} 
				& \multicolumn{2}{c}{\textbf{Toxicity}} 
				& \multicolumn{2}{c}{\textbf{FWER(\%)}} 
				& \multicolumn{2}{c}{\textbf{PCS(\%)}} 
				& \multicolumn{2}{c}{\textbf{Sample Size}} \\
				\cmidrule(lr){2-3}
				\cmidrule(lr){4-5}
				\cmidrule(lr){6-7}
				\cmidrule(lr){8-9}
				\cmidrule(lr){10-11}
				\textbf{Scenario} 
				& $q^{(true)}_{E,0}$& $q^{(true)}_{E,1}$
			& $q^{(true)}_{T,0}$ & $q^{(true)}_{T,1}$
				& BMW & Conv.
			& BMW & Conv.
			& BMW & Conv.\\
				\midrule
				
				1.1$^{\dagger}$ 
				& \multirow{4}{*}{(0.40, 0.30)} & (0.40, 0.66) 
				& \multirow{4}{*}{0.30} & 0.20
				& NA & NA
				& 41.6 & 43.4
				& 129.5 & 160.0 \\
				1.2$^{\ddagger}$
				&                          & (0.40, 0.66) 
				&                          & 0.30
				& 9.2 & 7.9
				& 90.8 & 92.1
				& 122.3 & 160.0 \\
				1.3$^{\mathsection}$
				&                          & (0.40, 0.30) 
				&                          & 0.20
				& 10.0 & 9.5
				& 94.6 & 95.5
				& 108.9 & 160.0 \\
				1.4$^{\mathparagraph}$
				&                          & (0.40, 0.30) 
				&                          & 0.30
				& 9.6 & 9.9
				& 99.6 & 99.6
				& 106.6 & 160.0 \\
				\midrule
				
				2.1$^{\dagger}$ 
				& \multirow{4}{*}{(0.40, 0.30)} & (0.50, 0.48) 
				& \multirow{4}{*}{0.30} & 0.20
				& NA & NA
				& 39.0 & 45.0
				& 131.0 & 160.0 \\
				2.2$^{\ddagger}$
				&                          & (0.50, 0.48) 
				&                          & 0.30
				& 5.3 & 6.0
				& 94.7 & 94.0
				& 122.7 & 160.0 \\
				2.3$^{\mathsection}$
				&                          & (0.50, 0.11) 
				&                          & 0.20
				& 9.7 & 9.5
				& 95.5 & 95.6
				& 107.6 & 160.0 \\
				2.4$^{\mathparagraph}$
				&                          & (0.50, 0.11) 
				&                          & 0.30
				& 10.3 & 10.1
				& 99.4 & 99.0
				& 106.2 & 160.0 \\
				\midrule
				
				3.1$^{\dagger}$ 
				& \multirow{4}{*}{(0.45, 0.35)} & (0.45, 0.73) 
				& \multirow{4}{*}{0.35} & 0.25
				& NA & NA
				& 40.1 & 44.0
				& 127.4 & 160.0 \\
				3.2$^{\ddagger}$
				&                          & (0.45, 0.73) 
				&                          & 0.35
				& 7.8 & 9.7
				& 92.2 & 90.3
				& 121.4 & 160.0 \\
				3.3$^{\mathsection}$
				&                          & (0.45, 0.35) 
				&                          & 0.25
				& 8.7 & 10.1
				& 96.4 & 95.9
				& 107.4 & 160.0 \\
				3.4$^{\mathparagraph}$
				&                          & (0.45, 0.35) 
				&                          & 0.35
				& 10.3 & 9.7
				& 99.2 & 99.6
				& 106.9 & 160.0 \\
				\midrule
				
				4.1$^{\dagger}$ 
				& \multirow{4}{*}{(0.45, 0.35)} & (0.55, 0.54) 
				& \multirow{4}{*}{0.35} & 0.25
				& NA & NA
				& 36.8 & 42.4
				& 132.2 & 160.0 \\
				4.2$^{\ddagger}$
				&                          & (0.55, 0.54) 
				&                          & 0.35
				& 9.0 & 8.9
				& 91.0 & 91.1
				& 124.0 & 160.0 \\
				4.3$^{\mathsection}$
				&                          & (0.55, 0.16) 
				&                          & 0.25
				& 7.7 & 10.3
				& 96.7 & 95.7
				& 109.4 & 160.0 \\
				4.4$^{\mathparagraph}$
				&                          & (0.55, 0.16) 
				&                          & 0.35
				& 10.2 & 9.7
				& 99.8 & 99.7
				& 106.6 & 160.0 \\
				\midrule
				
				5.1$^{\dagger}$ 
				& \multirow{4}{*}{(0.50, 0.40)} & (0.50, 0.78) 
				& \multirow{4}{*}{0.35} & 0.25
				& NA & NA
				& 40.9 & 39.9
				& 129.3 & 160.0 \\
				5.2$^{\ddagger}$
				&                          & (0.50, 0.78) 
				&                          & 0.35
				& 6.2 & 8.1
				& 93.8 & 91.9
				& 122.4 & 160.0 \\
				5.3$^{\mathsection}$
				&                          & (0.50, 0.40) 
				&                          & 0.25
				& 9.0 & 9.7
				& 96.1 & 96.2
				& 109.8 & 160.0 \\
				5.4$^{\mathparagraph}$
				&                          & (0.50, 0.40) 
				&                          & 0.35
				& 9.2 & 8.9
				& 99.8 & 99.4
				& 109.5 & 160.0 \\
				\midrule
				
				6.1$^{\dagger}$ 
				& \multirow{4}{*}{(0.50, 0.40)} & (0.60, 0.58) 
				& \multirow{4}{*}{0.35} & 0.25
				& NA & NA
				& 38.9 & 41.9
				& 130.2 & 160.0 \\
				6.2$^{\ddagger}$
				&                          & (0.60, 0.58) 
				&                          & 0.35
				& 7.2 & 7.7
				& 92.8 & 92.3
				& 122.2 & 160.0 \\
				6.3$^{\mathsection}$
				&                          & (0.60, 0.22) 
				&                          & 0.25
				& 7.5 & 9.8
				& 96.6 & 95.3
				& 105.6 & 160.0 \\
				6.4$^{\mathparagraph}$
				&                          & (0.60, 0.22) 
				&                          & 0.35
				& 9.8 & 9.8
				& 99.9 & 99.3
				& 106.9 & 160.0 \\
				\midrule
				
				7.1$^{\dagger}$ 
				& \multirow{4}{*}{(0.55, 0.45)} & (0.55, 0.82) 
				& \multirow{4}{*}{0.40} & 0.30
				& NA & NA
				& 37.3 & 42.2
				& 129.2 & 160.0 \\
				7.2$^{\ddagger}$
				&                          & (0.55, 0.82) 
				&                          & 0.40
				& 6.8 & 8.2
				& 93.2 & 91.8
				& 125.1 & 160.0 \\
				7.3$^{\mathsection}$
				&                          & (0.55, 0.45) 
				&                          & 0.30
				& 7.8 & 9.7
				& 96.5 & 95.6
				& 111.2 & 160.0 \\
				7.4$^{\mathparagraph}$
				&                          & (0.55, 0.45) 
				&                          & 0.40
				& 9.3 & 9.7
				& 99.5 & 99.8
				& 109.6 & 160.0 \\
				\midrule
				
				8.1$^{\dagger}$ 
				& \multirow{4}{*}{(0.55, 0.45)} & (0.65, 0.62) 
				& \multirow{4}{*}{0.40} & 0.30
				& NA & NA
				& 34.7 & 40.9
				& 130.2 & 160.0 \\
				8.2$^{\ddagger}$
				&                          & (0.65, 0.62) 
				&                          & 0.40
				& 6.4 & 8.1
				& 93.6 & 91.9
				& 123.7 & 160.0 \\
				8.3$^{\mathsection}$
				&                          & (0.65, 0.27) 
				&                          & 0.30
				& 7.6 & 9.5
				& 96.1 & 96.2
				& 108.3 & 160.0 \\
				8.4$^{\mathparagraph}$
				&                          & (0.65, 0.27) 
				&                          & 0.40
				& 9.5 & 10.9
				& 99.4 & 99.0
				& 108.2 & 160.0 \\
				\midrule
				
				9.1$^{\dagger}$ 
				& \multirow{4}{*}{(0.60, 0.50)} & (0.60, 0.85) 
				& \multirow{4}{*}{0.40} & 0.30
				& NA & NA
				& 38.1 & 41.3
				& 128.8 & 160.0 \\
				9.2$^{\ddagger}$
				&                          & (0.60, 0.85) 
				&                          & 0.40
				& 5.9 & 8.5
				& 94.1 & 91.5
				& 123.6 & 160.0 \\
				9.3$^{\mathsection}$
				&                          & (0.60, 0.50) 
				&                          & 0.30
				& 9.2 & 9.0
				& 96.2 & 95.7
				& 109.8 & 160.0 \\
				9.4$^{\mathparagraph}$
				&                          & (0.60, 0.50) 
				&                          & 0.40
				& 9.0 & 9.1
				& 99.7 & 99.2
				& 107.9 & 160.0 \\
				\midrule
				
				10.1$^{\dagger}$ 
				& \multirow{4}{*}{(0.60, 0.50)} & (0.70, 0.66) 
				& \multirow{4}{*}{0.40} & 0.30
				& NA & NA
				& 36.9 & 40.6
				& 129.7 & 160.0 \\
				10.2$^{\ddagger}$
				&                           & (0.70, 0.66) 
				&                           & 0.40
				& 6.6 & 7.9
				& 93.4 & 92.1
				& 123.8 & 160.0 \\
				10.3$^{\mathsection}$
				&                           & (0.70, 0.33) 
				&                           & 0.30
				& 9.6 & 8.1
				& 96.6 & 96.5
				& 110.6 & 160.0 \\
				10.4$^{\mathparagraph}$
				&                           & (0.70, 0.33) 
				&                           & 0.40
				& 9.5 & 8.9
				& 99.4 & 99.5
				& 108.3 & 160.0 \\
				\bottomrule
			\end{tabular}
		\end{adjustbox}
		
		\begin{tablenotes}[flushleft]
			{\footnotesize
			\item $^{\dagger}:$ $(\text{Efficacy pass},\,\text{Toxicity pass})$.
			\item $^{\ddagger}:$ $(\text{Efficacy pass},\,\text{Toxicity fail})$.
			\item $^{\mathsection}:$ $(\text{Efficacy fail},\,\text{Toxicity pass})$.
			\item $^{\mathparagraph}:$ $(\text{Efficacy fail},\,\text{Toxicity fail})$.
		}
		\end{tablenotes}
		
	\end{threeparttable}
\end{table}